\documentclass[runningheads]{llncs}

\usepackage{graphicx}
\usepackage[utf8]{inputenc}
\usepackage{tabularx}

\begin{document}
\title{AMUSED: An Annotation Framework of Multi-modal Social Media Data}
\titlerunning{AMUSED Annotation Framework}

\author{Gautam Kishore Shahi\inst{1}
\and 
Tim A. Majchrzak\inst{2}
}

\authorrunning{Shahi and Majchrzak}

\institute{University of Duisburg-Essen, Germany \and
University of Agder, Norway\\
\email{{gautam.shahi@uni-due.de}, \email{timam@uia.no}}
}
\maketitle 

\begin{abstract}
Social media nowadays is both an important news source and used for spreading misinformation.
Systematically studying social media phenomena, however, has been challenging due to the lack of labelled data. 
This paper presents the semi-automated annotation framework AMUSED for gathering multi-lingual multi-modal annotated data from social networking sites. The framework is designed to mitigate the workload in collecting and annotating social media data by cohesively combining machine and human in the data collection process. From a given list of news articles, AMUSED detects links to social media posts and then downloads the data from the respective social networking sites and assigns a label to it.  The framework can fetch the annotated data from multiple platforms like Twitter, YouTube, and Reddit. As a use case, we have implemented AMUSED for collecting COVID-19 misinformation data from different social media sites from 8\,077 fact-checked articles into four different categories of Misinformation.
\end{abstract}

\section{Introduction}
With the growth of users on different social media sites, social media have become part of our
lives. They play an essential role in making communication easier and accessible. People and organisations use social media to share and browse information, especially during the current pandemic; social media sites get massive attention from users \cite{talwar2019people,mcgahan2021secondary}.
Braun and Tarleton \cite{braun2011hosting} conducted a study to analyse the public discourse on social media sites and news organisation. Social media sites allow getting more attention from the users for sharing news or user-generated content. Several statistical or computational study has been conducted using social media data \cite{braun2011hosting}. But data gathering and its annotation are challenging and financially costly \cite{haertel2013practical}.

Social media data analytics research poses challenges in data collection, data sampling, data annotation, quality of the data, and bias in data \cite{grant2014enhancing}. Data annotation is the process of assigning a category to the data. Researchers annotate social media data for research on hate speech, misinformation, online mental health etc. For supervised machine learning, labelled data sets are required to understand the input patterns\cite{shahi2018analysis}.
To build a supervised or semi-supervised model on social media data, researchers face two challenges- timely data collection and data annotation \cite{shu2017fake}. Timely data collection is essential because some platforms either restrict data access or the post itself is deleted by social media platforms or by the user \cite{stieglitz2018social}. 
Another problem stands with data annotation; it is conducted either in an in-house fashion (within lab or organisation) or by using a crowdsourced tool (like Amazon Mechanical Turk (AMT)) \cite{aroyo2015truth}. Both approaches require a fair amount of effort to write the annotation guidelines along.
There is also a chance of wrongly labelled data leading to bias~\cite{cook2010automatically}.

We propose a semi-automatic framework for data annotation from social media platforms to solve timely data collection and annotation. AMUSED gathers labelled data from different social media platform in multiple formats (text, image, video).
It can get annotated data on social issues like misinformation, hate speech or other critical social scenarios.
AMUSED resolves bias in the data (wrong label assigned by annotator).
Our contribution is to provide a semi-automatic approach for collecting labelled data from different social media sites in multiple languages and data formats. Our framework can be applied in many application domains for which it typically is hard to gather the data, for instance, misinformation, mob lynching etc.

This paper is structured as follows.
In Section~\ref{sec:Related Work} we discuss the background of our work.
We then present the work method of AMUSED in Section~\ref{sec:method}, In Section~\ref{sec:implement} we give details on the implementation of AMUSED based on a case study. We discuss our observations in Section~\ref{sec:Discussion} and draw a conclusion in Section~\ref{sec:Conclusion}.

\section{Background}\label{sec:Related Work}
The following section describes the background on data annotation, types of data on social media, and the problem of the current annotations technique.

\subsection{Data Annotation}
Much research has been published that uses social media data.
Typically, it is limited to a few social media platforms or language in a single work. Also, the result is published with a limited amount of data. There are multiple reasons for these limitations; one of the key reason is the availability of annotated data for the research~\cite{thorson2013youtube,ahmed2013key}. Chapman et al.~\cite{chapman2011overcoming} highlight the problem of getting labelled data for an NLP related problem. A study is conducted on data quality and the role of annotator in the performance of machine learning model. With poor data, it hard to build a generalisable classifier~\cite{geiger2020garbage}.

Researchers are dependent on in-house or crowd-based data annotation. Recently, Alam et al. \cite{alam2020fighting} used a crowd-based annotation technique and asks people to volunteer for data annotation, but there is no significant success in getting a large number of labelled data. The current annotation technique is dependent on the background expertise of the annotators. 
Finding past data on an incident like mob lynching is challenging because of data restrictions by social media platforms. It requires looking at a massive number of posts and news articles, leading to much manual work. In addition, billions of social media posts are sampled to a few thousand posts for data annotation either by random sample or keyword sampling, leading to sampling bias.

With in-house data annotation, it is challenging to hire an annotator with background expertise in a domain. Another issue is the development of a \emph{codebook} with a proper explanation \cite{forbush2013catch}. The entire process is financially costly and time-taking \cite{duchenne2009automatic}. The problem with crowd-based annotation tools like AMT is that the low cost may result in the wrong labelling of data. Many annotators may cheat, not properly performing the job, use robots, or answer randomly~\cite{fort2011amazon,sabou2014corpus}.

Since the emergence of social media as a news resource~\cite{caumont201312}, people use this resource very differently. They may share news, state a personal opinion or commit a social crime in the form of hate speech or cyberbullying \cite{DBLP:conf/fire/0001MSJNPMS20}.
The COVID-19 pandemic arguably has to lead to a surge in the spread of misinformation~\cite{shahifakecovid}
Nowadays, journalists cover some common issues like misinformation, mob lynching, and hate speech; they also link the social media post in the news articles~\cite{cui2017does}.

To solve the problem of the data collection and its annotation, related social media posts from news articles can be used.
Labelling social media is then done based on the news article's contents. To get a reliable label, the credibility of the news sources must be considered~\cite{kohring2007trust}. For example, a professional news website registered with the International Fact-Checking Network~\cite{Poynter2020} should, generally, be rather creditable.

\subsection{Data on Social Media Platforms}
Social Media sites allow users to create and view posts in multiple formats. Every day, billions of posts containing images, text, videos are shared on social media sites such as Facebook, Twitter, YouTube and Instagram \cite{aggarwal2011introduction}. Data are available in different formats, and each social media platform apply restriction on data crawling. For instance,  Facebook allows crawling data only related to public posts and groups.

Giglietto et al. discuss the requirement of multi-modal data for the study of social phenomenon \cite{giglietto2012open}.
Almost every social media platform allows user to create or respond to the social media post in \emph{text}. But each social media platform has a different restriction on the length of the text. The content and the writing style changes with the character limit of different social media platform.
\emph{Images} are also common across different social media platforms. Platform have restriction on the size of the image.
Some platforms are primarily focused on \emph{video}, whereas some are multi-modal. Furthermore, for video, there are restrictions in terms of duration. This influences the characteristics of usage.

\subsection{Problems of Current Annotation Techniques}
There are several problems with the current annotation approaches. 
First, social media platforms restrict users when fetching data; for example, a user delete the tweets or videos on YouTube. Without on-time crawling, data access is lost.
Second, if the volume of data is high, filtering based on several criteria like keyword, date, location etc., is needed. This filtering degrades the data quality by excluding much data. For example, if we sample data using hateful keywords for hate speech, we might lose many hate speech tweets but do not contain any hateful words.

Third, getting a good annotator is a difficult task. Annotation quality depends on the background expertise of the person. For crowdsourcing, maintaining annotation quality is complicated. Moreover, maintaining a good agreement between multiple annotators is tedious.
Fourth, the development of annotation guidelines is tricky. Writing a good codebook requires domain knowledge and consultation from experts.

Fifth, data annotation is costly and time-consuming \cite{sorokin2008utility}.
Sixth, social media is available in multiple languages, but much research is limited to English. Data annotation in other languages, especially under-resourced languages, is difficult due to the lack of experienced annotators.

\section{Method}\label{sec:method}
\begin{figure}[t!]
    \centering
    \includegraphics[width=1\textwidth]{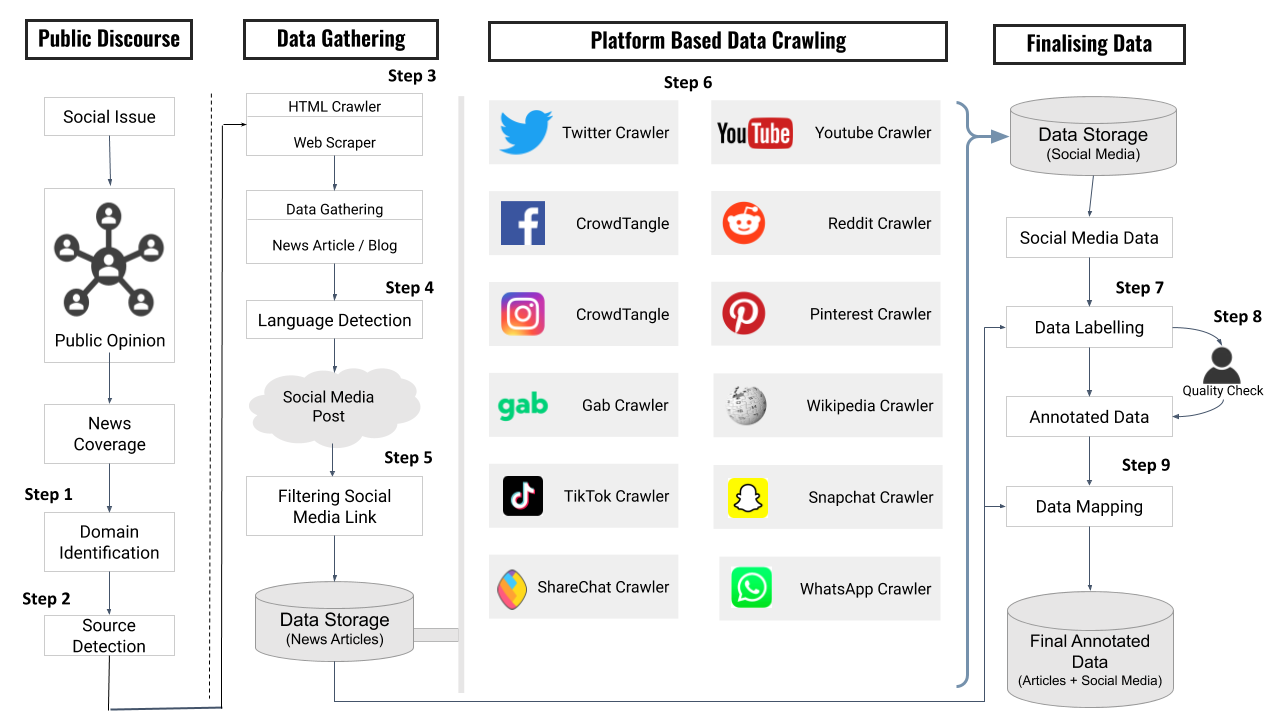}
    \caption{AMUSED: An Annotation Framework for Multi-modal Social Media data}
    \label{figure:framework}
    \vspace{-1em}
\end{figure}

AMUSED's elements are summarised in Figure~\ref{figure:framework}.
It follows nine steps.

\textbf{Step 1: Domain Identification} The first step is the identification of the domain in which we want to gather the data. A domain could focus on a particular public discourse. For example, a domain could be \emph{fake news in the US election}, or \emph{hate speech in trending hashtags on Twitter}. Domain selection helps to find the relevant data sources.

\textbf{Step 2: Data Source} Data sources comprise news websites that mention a particular topic. For example, many news websites have a separate section that discusses the election or other ongoing issues. 

\textbf{Step 3: Web scraping} AMUSED then crawls all news articles from news websites using a Python-based crawler. We fetch details such as the published date, author, location, news content (see Table~\ref{tab:element}).

\begin{table}[t!]
\centering
\begin{tabularx}{\linewidth}{p{2.5cm}X}
    \textbf{Element} &    \textbf{Definition}   \\ \hline
   News\_ID & Unique identifying ID of each news articles. We use an acronym for news source and the number to identify a news articles. \\
   & Example: \textit{PY9} \\ 
 Newssource\_URL & Unique identifier pointing to the news articles.  \\
 & Example: \textit{\url{https://factcheck.afp.com/video-actually-shows-anti-government-protest-belarus}}\\ 
News\_Title & The title of the news article.   \\
&  Example: \textit{A video shows a rally against coronavirus restrictions in the British capital of London.}\\ 
Published\_date & Date when an article published in online media. \\
& Example: \textit{01 September 2020} \\ 
News\_Class & Each news articles published the fact check article with a class like false, true, misleading. We store it in the class column. \\
& Example: \textit{False}\\
  Published-By & The name of the news websites\\
& Example: \textit{AFP, TheQuint}\\
Country & Country where the news article is published. \\
& Example: \textit{Australia}\\ 
Language &  Language used for news article. \\
& Example: \textit{English}\\ 
\end{tabularx}
\caption{Description of attributes and their examples}
\label{tab:element}
\vspace{-2em}
\end{table}

\textbf{Step 4: Language Identification} After getting the details from the news articles, we check its language. We use ISO 639-1 for naming the language. Based on the language, we can further filter articles and apply a language-specific model for finding insights. 

\textbf{Step 5: Social Media Link} From the crawled data, we fetch the anchor tag \texttt{\textless{}a\textgreater{}} mentioned in the news content.
We then filter the hyperlinks to identify social media platforms and fetch unique identifiers to the posts.

\textbf{Step 6: Social Media Data Crawling} We now fetch the data from the respective social media platform. For this purpose, we built a crawler for each social media platform, which consumes the unique identifiers obtained from the previous step.
For \emph{Twitter} we used a Python crawler using Tweepy, which crawls all details about a Tweet. We collect text, time, likes, retweet, user details such as name, location, follower count.
Similarly, we build our crawler for other platforms. Due to the data restriction from Facebook and Instagram, we use Crowdtangle \cite{team2020crowdtangle} to fetch data from Facebook and Instagram, but it only gives numerical data like likes and followers.

\textbf{Step 7: Data Labelling} We assign labels to the social media data based on the label assigned to the news articles by journalists. Often news articles categorise a social media post, for example, like hate speech or propaganda. We assign the label to social media post as class mentioned in the news article as a class described by the journalist. For example, suppose a news article $a$ containing social media post $s$ has been published by a journalist $j$, and journalist $j$ has described the social media post $s$ to be misinformation. In that case, We label the social media post $s$ as misinformation. It will ease the workload by getting the number of social media post check by a journalist.

\textbf{Step 8: Human Verification} To check the correctness, a human verifies the assigned label to the social media post.
If the label is wrongly assigned, then data is removed from the corpus. This step assures that the collected social media post contains the relevant post and correctly given label. A human can verify the label of the randomly selected news articles.

\textbf{Step 9: Data Enrichment} We finally merge the social media data with the details from the news articles. It helps to accumulate extra information, which might allow for further analysis.

\section{Implementation: A Case Study on Misinformation}
\label{sec:implement}
While our framework allows for general application, understanding its merits is best possible by applying it to a specific domain. AMUSED can be helpful for several domains, but news companies are quite active in the domain of misinformation, especially during a crisis.
\emph{Misinformation}, often yet imprecisely referred to as a piece of information that is shared unintentionally or by mistake, without knowing the truthfulness of the content \cite{shahi2021exploratory}.

There is an increasing amount of Misinformation in the media, social media, and other web sources; this has become a topic of much research attention~\cite{abs-1812-00315}. Nowadays, more than 100 fact-checking websites are working to tackle the problem of misinformation \cite{cherubini2016rise}.

People have spread vast amounts of misinformation during the COVID-19 pandemic and in relation to elections and disasters~\cite{gupta2013faking}. Due to the lack of labelled data, it is challenging to make a proper analysis of the misinformation.

As a case study, we apply the AMUSED for data annotation for COVID-19 misinformation, following the steps illustrated in the prior section. 

\textbf{Step 1: Domain Identification} Out of several possible application domains, we consider the spread of misinformation in the context of COVID-19. Misinformation likely worsens the negative effects of the pandemic~\cite{shahifakecovid}.
The director of the World Health Organization (WHO) considers that we are not only fighting with a pandemic but also an \emph{infodemic} \cite{GuardianWHO,Zarocostas2020}.
One of the fundamental problems is the lack of sufficient corpus related to pandemic \cite{shahi2021exploratory}.

\textbf{Step 2: Data Sources} For data source, we analysed 25 fact-checking websites and decided to use  \emph{Poynter} and \emph{Snopes}. We choose Poynter because it has a central data hub that collects data from more than 98 fact-checking websites, while Snopes is not integrated with Poynter but has more than 300 fact-checked articles on COVID-19.

\textbf{Step 3: Web Scraping}
In this step, we fetched all the news articles from Poynter and Snopes.

\textbf{Step 4: Language Detection} We collected data in multiple languages like English, German, and Hindi. To identify the language of the news article, we have used langdetect,
a Python-based library to detect the language of the news articles. We used the textual content of new articles to check the language of the news articles.

\begin{figure}[thb]
    \centering
    \vspace{-1.8em}
    \includegraphics[width=\linewidth,height=7cm]{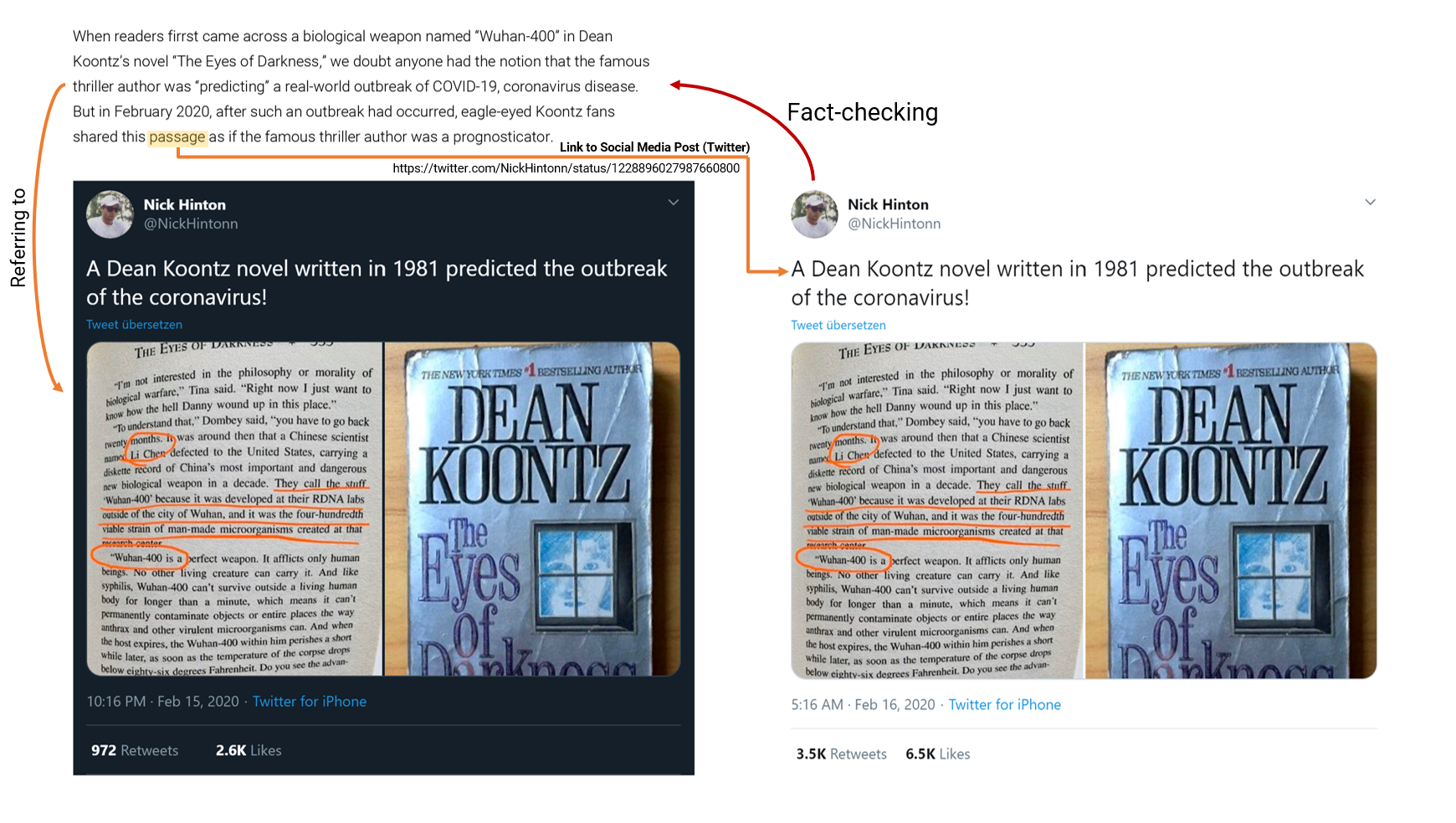}
    \vspace{-2.5em}
    \caption{An Illustration of data collection from social media platform(Twitter) from a news article \cite{shahi2021exploratory}}
    \label{figure:news}
    \vspace{-1.0em}
\end{figure}

\textbf{Step 5: Social Media Link}
In the next step, while doing HTML crawling, we filter the URL from the parsed tree of the DOM (Document Object Model). We analysed the URL pattern from different social media platforms and applied keyword-based filtering from all hyperlinks in the DOM. For instance, For each Tweet, Twitter follows a pattern twitter.com/user\_name/status/tweetid. So, in the collection hyperlink, we searched for the keyword ``twitter.com'' and ``status''. This assures that we have collected the hyperlink referring to the tweet.
This process is shown in Figure~\ref{figure:news}.  

Similarly, we followed the approach for other social media platforms like Facebook and Instagram. We used the regex code to filter the unique ID for each social media post in the next step.

\textbf{Step 6: Social Media Data Crawling}
We now have the unique identifier of each social media post. We built a Python-based program for crawling the data from the respective social media platform. The summary is given in Table~\ref{tab:datasum}.

\begin{table}[thb]
\centering
\begin{tabular}{l|r|r|r|r|r|r}
    \textbf{Platform}&    \textbf{Posts} &  \textbf{Unique} & \textbf{Text} & \textbf{Image} & \textbf{Text+Image} & \textbf{Video} \\
    Facebook      & 5\,799 & 3\,200  & 1167 & 567 & 1\,006 & 460\\
    Instagram    & 385 & 197  & - & 106 & 41  &52\\
    Pinterest       & 5 & 3 & -  & 3 & 0 & 0\\
    Reddit      & 67 & 33 & 16 & 10 & 7 & 0\\
    TikTok     & 43 & 18 & - & -& -&18\\
    Twitter  & 3\,142 & 1\,758 & 1300 &116 &143 &199\\
    Wikipedia        & 393 & 176 & 106  & 34& 20&16\\
    YouTube      & 2\,087~(916) & -  &- & - & 916\\
\end{tabular}
\caption{Summary of data collected}
\label{tab:datasum}
\vspace{-1em}
\end{table}

\begin{figure}[thb]
    \centering
    \includegraphics[width=0.7\textwidth]{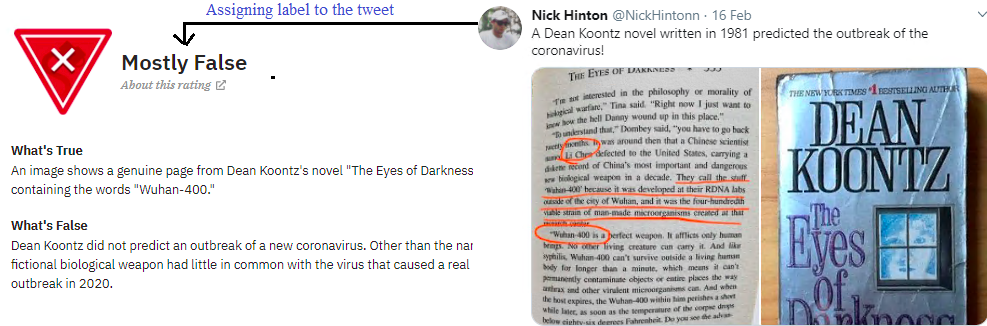}
    \vspace{-1em}
    \caption{An Illustration for annotation of social media posting using the label mentioned in the news article.}
    \label{figure:dataannotation}
    \vspace{-1.0em}
\end{figure}

\textbf{Step 7: Data Labelling} For data labelling, we used the label assigned in the news articles, then we mapped the social media post with their respective news article and assigned the label to the social media post. For example, a Tweet extracted from a news article is mapped to the class of the news article. This process is shown in Figure~\ref{figure:dataannotation}.

\textbf{Step 8: Human Verification} We manually checked each social media post to assess the correctness of the process. We provided the annotator with all necessary information about the class mapping and asked them to verify it. For example, in Figure~\ref{figure:dataannotation}, a human open the news article using the newssource\_URL and verifies the label assigned to the tweet. For COVID-19 misinformation, we check the annotation by randomly choosing 100 social media posts from each social media platform and verifying the label assigned to the social media post and label mentioned in the news articles. We measured the inter-coder reliability using Cohen's kappa and got a value of 0.72-0.86, which is a good agreement.
We further normalised the data label into \emph{false}, \emph{partially false}, \emph{true} and \emph{others} using the definitions mentioned in~\cite{shahi2021exploratory}. 

\textbf{Step 9: Data Enrichment} We then enriched the data by providing extra information about the social media post. The first step is merging the social media post with the respective news article, and it includes additional information like textual content, news source, author.
The detailed analysis of the collected data is discussed in the result section.

\section{Results}
For the use case of COVID-19 Misinformation, we identified Poynter and Snopes as the data source, and we collected data from different social media platforms. We found that around 51\% of news articles linked their content to social media websites. Overall, we have collected 8,077 fact-checked news articles from 105 countries in 40 languages.
We have cleaned the hyperlinks collected using the AMUSED framework and filtered the social media posts by removing the duplicates using the unique identifier. Finally, we will release the data as open-source.

\begin{table}[t]
\centering
\begin{tabular}{l|r|r|r|r|r}
    \textbf{SM Platform}&    \textbf{False} & \textbf{Partially False} & \textbf{Other} & \textbf{True} \\
    Facebook    & 2,776 &  325 & 94 & 6 &\\
    Instagram    & 166  & 28 & 2 & 1 \\
    Reddit      &  21 & 9 & 2 & 1\\
    Twitter  & 1,318  & 234 & 50 & 13\\
    Wikipedia        & 154  & 18  & 3  & 1\\
    YouTube      &  739 & 164  & 13 & 0\\
\end{tabular}
\caption{Summary of COVID-19 misinformation posts collected.}
\label{tab:datafake}
\vspace{-2em}
\end{table}

We plotted the data from those social media platform which has the total number of post more than 25 unique posts in Table \ref{tab:datafake} because it depreciates the plot distribution. We dropped the plot from Pinterest~(3), Whatsapp~(23), Tiktok~(25), Reddit~(43).
The plot shows that most of the social media posts are from Facebook and Twitter, followed by YouTube, Wikipedia and Instagram. Table~\ref{tab:datafake} also presents the class distribution of these posts.
Misinformation also follows the COVID-19 situation in many countries because the number of social media posts also decreased after June 2020. The possible reason could be either that the spread of Misinformation is reduced or that fact-checking websites are not focusing on this issue as during the early stage.

\section{Discussion}\label{sec:Discussion}
Our study highlighted the process of fetching the labelled social media post from news fact-checked articles.
Usually, the fact-checking website links the social media post from multiple social media platforms. We tried to gather data from various social media platforms, but we found the maximum number of Facebook, Twitter, and YouTube links. There are few unique posts from Reddit~(21), TikTok~(9) etc., which shows that fact-checker mainly focused on analysing content from Facebook, Twitter, and YouTube. 

Surprisingly there are only three unique posts from Pinterest, and there are no data available from Gab, ShareChat, and Snapchat. However, Gab is well known for harmful content, and people in their regional languages use ShareChat.
There are only three unique posts from Pinterest. Many people use Wikipedia as a reliable source of information, but there are 393 links from Wikipedia. Hence, overall fact-checking website is limited to some trending social media platforms like Twitter or Facebook, while social media platforms like Gab, TikTok is famously famous for malformation or misinformation \cite{brennen2020types}. WhatsApp is an instant messaging app used among friends or group of people. So, we only found some hyperlink which links to the public WhatsApp group. To increase the visibility of fact-checked articles, a journalist can also use schema.org vocabulary along with the Microdata, RDFa, or JSON-LD formats to add details about Misinformation to the news articles \cite{shahi2019inducing}.

AMUSED requires some effort but still is beneficial compared to random data annotation because we need to annotate thousands of social media posts. Still, the chances of getting misinformation are significantly less.

Another aspect is the diversity of social media post on the different social media platforms. News articles often mention Facebook, Twitter, YouTube, yet only seldom Instagram, Pinterest, Gab and Tiktok were not mentioned at all. The reasons for this need to be explored. It would be interesting to study the propagation of misinformation on different platforms like Tiktok and Gab in relation to the news coverage they get.
Such a cross-platform study would particularly insightful with contemporary topics such as misinformation on COVID-19.
Such a cross-platform work could also be linked to classification models~\cite{shahi2018analysis,nandini2018modelling}.

We have also analysed the multi-modality of the data on the social media platform; the number of social media post is shown in Table~\ref{tab:datasum}. We further classify the misinformation into four different categories, as discussed in step 8. The amount of Misinformation as text is greater compared to video or image. Thus, in Table~\ref{tab:datafake} we present the textual misinformation into four different categories. Apart from text, the misinformation is also shared as image, video or embedding format like image-text.

While applying the AMUSED framework on the misinformation on COVID-19, we found that misinformation spreads across multiple source platforms, but it mainly circulated across Facebook, Twitter, YouTube. Our finding suggests concentrating mitigation efforts onto these platforms.

\section{Conclusion and Future Work}\label{sec:Conclusion}
In this paper, we presented a semi-automatic framework for social media data annotation. The framework can be applied to several domains like misinformation, mob lynching, and online abuse. As a part of the framework, we also used a Python-based crawler for different social media websites. After data labelling, the labels are cross-checked by a human, which ensures a two-step verification of data annotation for the social media posts. We also enrich the social media post by mapping it to the news article to gather more analysis about it. The data enrichment will be able to provide additional information for the social media post. We have implemented the proposed framework for collecting the misinformation post related to the COVID-19. One of the limitations of the framework is that, presently, we do not address the multiple (possibly contradicting) labels assigned by different fact-checkers over the same claim.

As future work, the framework can be extended for getting the annotated data on other topics like hate speech, mob lynching etc. The framework will be helpful in gathering annotated data for other domains from multiple social media sites for further analysis. 

AMUSED will decrease the labour cost and time for the data annotation process. Our framework will also increase the data annotation quality because we crawl the data from news articles published by an expert journalist.

\bibliography{amused}
\bibliographystyle{splncs04}

\end{document}